\documentclass[letter]{aa} 

\usepackage{color}
\usepackage{graphicx}
\usepackage[fleqn]{amsmath}	
\usepackage{amssymb}	
\usepackage{mathtools}
\usepackage{txfonts}
\usepackage{booktabs}
\usepackage{natbib,twoopt}
\usepackage[T1]{fontenc}
\usepackage{float}
\usepackage[caption = false]{subfig}
\usepackage[breaklinks=true]{hyperref}
\hypersetup{
  colorlinks=true,   
  citecolor=blue,    
  urlcolor=blue,     
  linkcolor=blue,     
}

\bibpunct{(}{)}{;}{a}{}{,} 
\usepackage{lscape}

\DeclareRobustCommand{\ion}[2]{%
\relax\ifmmode
\ifx\testbx\f@series
{\mathbf{#1\,\mathsc{#2}}}\else
{\mathrm{#1\,\mathsc{#2}}}\fi
\else\textup{#1\,{\mdseries\textsc{#2}}}%
\fi}

\newcommand{\mbh}{$M_{\rm BH}$}
\newcommand{\ha}{H$\alpha$}
\newcommand{\hb}{H$\beta$}

\def \arcsec    {$^{\prime\prime}$}

\newcommand{\xmm}{XMM--\emph{Newton}\,}

\begin{document}

   \title{The optically elusive, changing-look active nucleus in NGC 4156}

   \author{Giulia Tozzi\inst{1,2}\thanks{\email{giulia.tozzi@unifi.it}},
    Elisabeta Lusso\inst{1,2},
    Lapo Casetti\inst{1,2,3},
    Marco Romoli\inst{1,2},
    Gloria Andreuzzi\inst{4,5},
    Isabel  Montoya A.\inst{6},
    Emanuele Nardini\inst{2},
    Giovanni Cresci\inst{2},
    Riccardo Middei\inst{5,7},
    Silvia Bertolini\inst{1},
    Paolo Calabretto\inst{8},
    Vieri Cammelli\inst{9,10},
    Francisco Cuadra\inst{1},
    Marco Dalla Ragione\inst{11},
    Cosimo Marconcini\inst{1},
    Adriano Miceli\inst{1},
    Irene Mini\inst{12},
    Martina Palazzini\inst{1},
    Giorgio Rotellini\inst{1},
    Andrea Saccardi\inst{13},
    Lavinia Sam\`{a}\inst{14},
    Mattia Sangalli\inst{1},
    Lorenzo Serafini\inst{1},
    Fabio Spaccino\inst{1}
    }

   \institute{$^{1}$Dipartimento di Fisica e Astronomia, Università di Firenze, Via G.\ Sansone 1, 50019 Sesto Fiorentino, FI, Italy\\
$^{2}$INAF - Osservatorio Astrofisico di Arcetri, Largo E.\ Fermi 5, 50125 Firenze, Italy\\
$^{3}$INFN - Sezione di Firenze, Via G.\ Sansone 1, 50019 Sesto Fiorentino, FI, Italy\\
$^{4}$Fundaci\'{o}n Galileo Galilei, Rambla Jos\'{e} Ana Fernandez P\'{e}rez 7, 38712 Bre\~{n}a Baja, TF, Spain\\
$^{5}$INAF - Osservatorio Astronomico di Roma, via Frascati 33, 00078 Monte Porzio Catone, RM, Italy\\
$^{6}$Institute of Theoretical Astrophysics, University of Oslo, Postboks 1029 Blindern 0315, Oslo, Norway\\
$^{7}$Space Science Data Center (SSDC), ASI, Via del Politecnico snc, 00133 Roma, Italy\\
$^{8}$INAF – Osservatorio di Astrofisica e Scienza dello Spazio di Bologna, Via Gobetti 93/3, 40129 Bologna, Italy\\
$^{9}$Dipartimento di Fisica, Università di Trieste, Via A.\ Valerio 2, 34127 Trieste, Italy\\
$^{10}$INAF - Osservatorio Astronomico di Trieste, Via Tiepolo 11, 34143 Trieste, Italy\\
$^{11}$Dipartimento di Matematica e Informatica ``Ulisse Dini'', Università di Firenze, Viale G.\ B.\ Morgagni 67/a, 50134 Firenze, Italy\\
$^{12}$Dipartimento di Fisica e Astronomia ``Augusto Righi'', Università di Bologna, Via P.\ Gobetti 93/2, 40129 Bologna, Italy\\
$^{13}$GEPI, Observatoire de Paris, Universit\'{e} PSL, CNRS, 5 Pl.\ Jules Janssen, 92190 Meudon, France\\
$^{14}$Dipartimento di Fisica ``E. Fermi'', Università di Pisa, Largo Bruno Pontecorvo 3, 56127 Pisa, Italy
             }

  \titlerunning{The changing-look active nucleus in NGC 4156}
  \authorrunning{G. Tozzi et al.}
   \date{\today}


  \abstract{We report on the changing-look nature of the active galactic nucleus (AGN) in the galaxy NGC 4156, as serendipitously discovered thanks to data acquired in 2019 at the Telescopio Nazionale Galileo (TNG) during a students' observing programme. Previous optical spectra had never shown any signatures of broad-line emission, and evidence of the AGN had come only from X-ray observations, being the optical narrow-line flux ratios unable to unambiguously denote this galaxy as a Seyfert. Our 2019 TNG data unexpectedly revealed the appearance of broad-line components in both the H$\alpha$ and H$\beta$ profiles, along with a rise of the continuum, thus implying a changing-look AGN transitioning from a type 2 (no broad-line emission) towards a (nearly) type 1. The broad-line emission has then been confirmed by our 2022 follow-up observations, whereas the rising continuum has no longer been detected, which hints at a further evolution backwards to a (nearly) type 2. The presence of broad-line components also allowed us to obtain the first single-epoch estimate of the black hole mass ($\log(M_{\rm BH}/\text{\(M_\odot\)})\sim8.1$) in this source. The observed spectral variability might be the result of a change in the accretion activity of NGC 4156, although variable absorption cannot be completely excluded.} 


   \maketitle
%

\section{Introduction}
Active galactic nuclei (AGN) represent a key stage of the galaxy life cycle, powered by the accretion of matter onto the supermassive black hole (SMBH; with a black hole mass $M_{\rm BH}\simeq10^6-$10$^{10}$ \(M_\odot\)) located at their core. This active phase is characterised by high luminosity and strong variability at all wavelengths. Depending on the presence or absence of broad optical emission lines (full width at half maximum FWHM $>2000$ km\,s$^{-1}$), active galaxies are commonly classified into type 1 or type 2, respectively. The two classes of AGN are historically explained in terms of a dusty and gaseous torus differently oriented with respect to the observer, which shields the broad line region (BLR) emission in type 2 AGN, but not in type 1 AGN \citep{1993ARA&A..31..473A, 2015ARA&A..53..365N}. Intermediate types can then result from the combination of viewing angles and a clumpy distribution of the obscuring material with different optical depths \citep{1989ApJ...340..190G}.

In recent years, a number of AGN, the so-called Changing Look AGN (CLAGN), have been discovered undergoing type transitions within a few years or even several months.
 In the optical, changing-look phenomena are typically identified through the appearance (disappearance) of the broad \ha\ and \hb\ components in previously known type 2 (type 1) AGN (e.g. \citealt{2014ApJ...796..134D, 2014ApJ...788...48S, 2019MNRAS.486..123R}).
 The local Seyfert NGC 1566 is one of the first CLAGN ever discovered \citep{1970ApL.....6..155P}. Its variability has been widely studied from the optical, to the UV and X-ray energies, revealing a dramatic variability of emission line profiles and recurrent brightening events \citep{1986ApJ...308...23A,2017MNRAS.470.3850D,2019MNRAS.483..558O,2022AN....34310080O}. The changing state of NGC 1566 was interpreted as the result of a luminosity increase causing the sublimation of dust in the line of sight, which previously obscured part of the BLR.
However, in spite of the increasing number of newly discovered CLAGN (a few tens), the physical mechanism at the origin of CLAGN is still under debate.

In this Letter, we report on the intriguing serendipitous discovery of the galaxy \object{NGC 4156} as a new CLAGN. NGC 4156 is a face-on, barred spiral galaxy (SBc; \citealt{1984A&A...131..291N}) at $z=0.0226$ (for a luminosity distance of 100 Mpc, \citealt{2015ApJS..219...12A}), whose AGN nature is implied by X-ray observations \citep{elvis1981, 2005A&A...444..119G}. Based on its optical narrow-line emission, NGC 4156 has been classified as a LINER (low-ionisation nuclear emission-line region; \citealt{1992ApJ...388..310K,2016MNRAS.455.2551N}) galaxy, with no definitive optical AGN signatures. In particular, the Sloan Digital Sky Survey (SDSS DR6; \citealt{2008ApJS..175..297A}) spectrum, taken in 2004, is devoid of both a bright AGN continuum and broad \ha\ and \hb\ components. From the stellar velocity dispersion measured in the SDSS data ($\sigma_*=155.2$\,km\,s$^{-1}$), the mass of the SMBH in NGC 4156 is estimated to be log$(M_{\rm BH}/\text{\(M_\odot\)})\approx 7.7$ \citep{2016MNRAS.455.2551N} via the $M_{\rm BH}-\sigma_*$ relation \citep{2013ApJ...764..184M}. Radio emission from NGC 4156 was also detected during the Very Large Array FIRST survey \citep{2004A&A...416...35W}, with a radio flux density at an observed frequency $\nu_{\text{obs}}=1.4$ GHz of 2.93 mJy, corresponding to a radio luminosity $L_{\text{1.4 GHz}}\approx 10^{37.7}$\,erg\,s$^{-1}$. The radio source is estimated to lie $1.04''$ apart from the AGN location \citep{2004A&A...416...35W}.

\citet{elvis1981} were the first to report the detection by the Einstein Observatory High Resolution Imager of a compact, likely variable, luminous X-ray source in the optically ``dull'' NGC 4156 galaxy. Such a strong X-ray activity ($L_{\rm X}\sim10^{42}$ erg s$^{-1}$) was indeed unexpected based on the optical spectrum acquired with the Multiple Mirror Telescope, displaying the weakest emission-line fluxes of any previously known X-ray AGN.


In 2019, NGC 4156 was observed by undergraduate students of the University of Florence (Italy) at the Telescopio Nazionale Galileo (TNG) located at the Roque de Los Muchachos (La Palma, Canary Islands). The optical spectrum acquired with the spectrograph Dolores (Device Optimized for the LOw RESolution) has given the first clues about the ongoing state transition of NGC 4156 from a type 2 towards a type 1 AGN (see Figure~\ref{fig:comparison}). Therefore, new TNG/Dolores observations of NGC 4156 have been recently performed (P.I.: E.\ Lusso), with longer exposure times and at higher spectral resolution, aimed at confirming the CLAGN nature of NGC 4156 and at robustly modelling the broad-line component of \ha\ and \hb\ to estimate the black hole mass and the accretion rate of the source. 
Here we present the TNG/Dolores observations obtained in 2019 and 2022, which together lead to the discovery of NGC 4156 undergoing a double changing-look transition: first, from a type 2 (in 2004) towards a type 1 (i.e. type 1.5 in 2019), and then backwards towards a type 2 (i.e. type 1.8 in 2022).

\section{TNG/Dolores observations of NGC 4156}
\label{sec:obs}
NGC 4156 was observed at the TNG on April 28$^{\rm th}$ 2019 during an observing campaign led by undergraduate students from the University of Florence (see Appendix~\ref{apx:obscampaign} for details).
The spectrum of NGC 4156 was acquired with Dolores using the LR-B grism, covering the wavelength range 3000--8400 \AA. NGC 4156 was observed together with the nearby NGC 4151, which was actually the primary target of the observation, by placing the $1^{\prime\prime}$ slit along the direction joining the two galaxies. The exposure time (300 s) was tailored to the latter, which is brighter 
than NGC 4156.
The seeing varied between 0.5\arcsec and 1.5\arcsec.
The 2019 data of NGC 4156 were reduced for the first time in 2022, hence revealing unexpected hints on its transition from a type-2 to a type-1 Seyfert. After such a discovery, we followed up the galaxy on May 28$^{\rm th}$ 2022 with Dolores using the 1.5\arcsec\ slit and two different spectral setups: (i) LR-B grism with exposure time of 1200 seconds; (ii) VHR-R grism (6300--7700 \AA) with exposure time of 1800 seconds. The average seeing was 1.5\arcsec. 

\subsection{Data reduction}
We reduced the LR-B TNG spectra taken in 2019 and 2022 with the open source data reduction pipeline \textsc{PypeIt} 
 \citep[][v1.9]{pypeit:joss_pub,pypeit:zenodo}. 
The data reduction follows the standard procedure, including bias subtraction and flat fielding, removal of cosmic rays, sky subtraction, extraction of the one-dimensional (1D) spectrum and wavelength calibration. The latter has been derived from the combined helium, neon and mercury arc-lamps in the vacuum frame, and a quadratic polynomial function was used to fit the pixel-wavelength data. The resulting dispersion from the wavelength calibration is $\Delta\lambda=2.72$ \AA/px for both spectra with an $\text{rms}< 0.1$ px and $\approx 0.15$ px for the 2019 and 2022 data sets, respectively. 
The nuclear integrated spectra were extracted using a pseudo-slit of $1.5'' \times 4''$ centered on the galaxy centre ($\text{PA} = 39^\circ$ and $\text{PA} = 0^\circ$ for the 2019 and 2022 data, respectively). 
A standard star with known spectral type was observed during both the 2019 (HD 121409) and 2022 (Feige 66) observing nights to account for telluric absorption and flux calibration, whose data were reduced with the same pipeline.

The reduction of the VHR-R data was performed with a custom-made pipeline, since this grism is not currently supported in \textsc{PypeIt}. The custom pipeline uses some routines from MAAT \citep[MATLAB Astronomy and Astrophysics Toolbox,][]{Ofek2014}. It includes all the standard corrections listed above for \textsc{PypeIt}. The wavelength calibration was performed as per the LR-B data. The resulting dispersion is $\Delta\lambda=0.78$ \AA/px.
The telluric correction and the flux calibration were performed with the \textsc{IDL} software \textsc{XTELLCORR} \citep{vacca2003}. 
The nuclear integrated spectrum was extracted using a pseudo-slit of $1.5^{\prime\prime} \times 4^{\prime\prime}$ centered on the galaxy center ($\text{PA} = 0^\circ$).
To test the robustness of our pipeline, we also reduced the 2019 and 2022 LR-B data, obtaining results fully consistent with the \textsc{PypeIt} output.

\begin{figure*}
\centering\includegraphics[width=0.8\hsize]{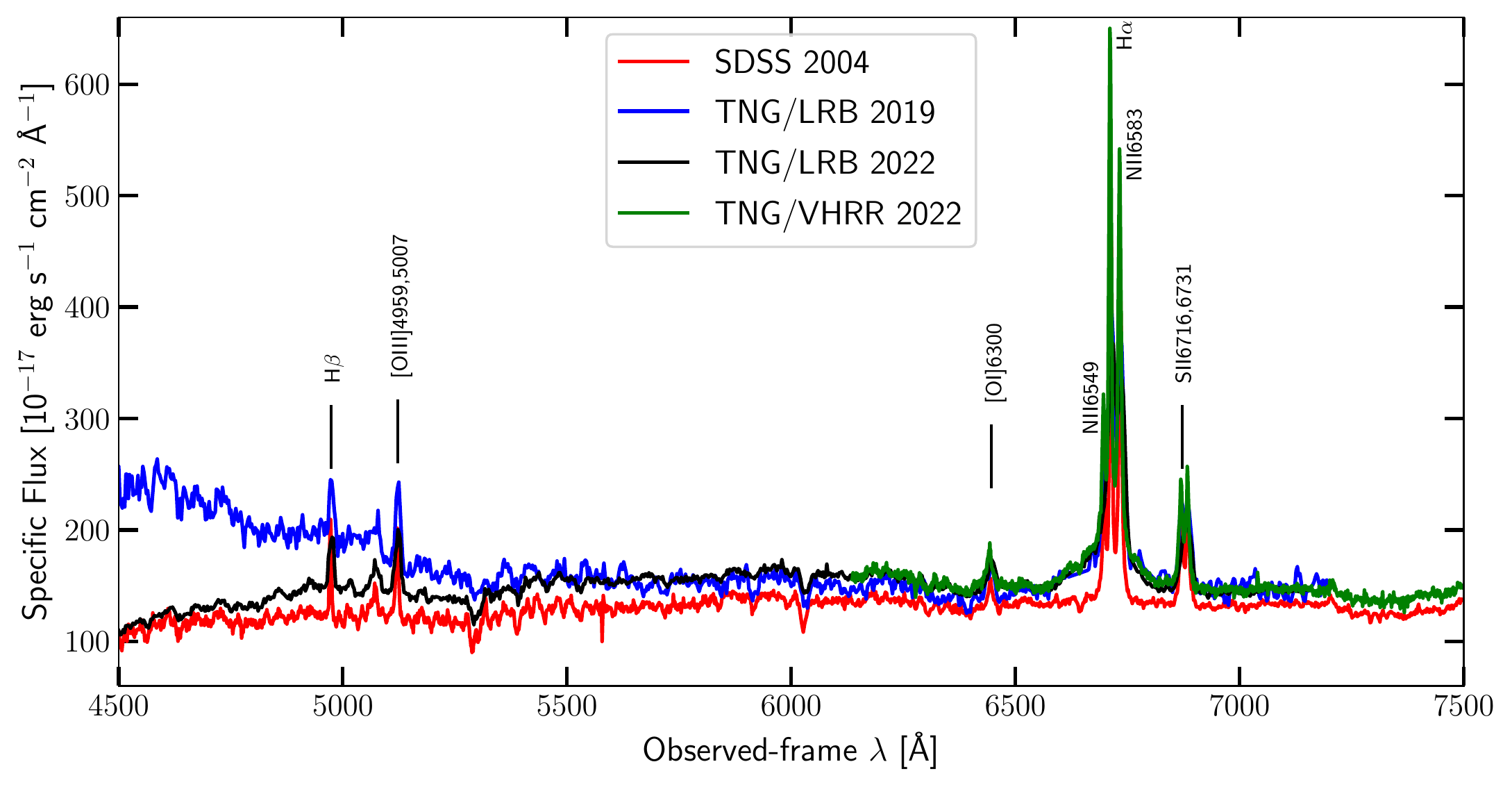} 
    \caption{Comparison of the SDSS 2004 spectrum (red) with the spectra acquired at the TNG. The TNG/LR-B data observed in 2019 and 2022 are represented by blue and black lines, respectively; while the TNG/VHR-R 2022 spectrum by a green line. Data are shown in the observed frame.}
    \label{fig:comparison}
\end{figure*}
The final 1D TNG LR-B and VHR-R observed spectra are shown in Figure~\ref{fig:comparison} together with the archival SDSS data. From a first visual comparison, the TNG/2019 spectrum clearly shows a broad \ha\ component (FWHM$\approx$4000 km s$^{-1}$) unlike the SDSS spectrum, and a similar broad component seems to be present for the \hb\ emission line as well.
Additionally, the TNG/2019 spectrum displays a rise of the continuum at $\lambda<5500$ \AA, likely due to thermal emission from the accretion disc around the SMBH. 
Interestingly, the rising continuum is no longer detected in the TNG/2022 data, whilst the broad \ha\ component is still clearly observed. We will discuss these results in Section~\ref{sec:discussion}.

\section{Spectral analysis}
\label{sec:spanalysis}
We fitted the four 1D spectra of NGC 4156 (i.e. the SDSS 2004, the TNG/LR-B 2019, and the TNG/LR-B and VHR-R 2022) with both the fitting code described in \citet[][see also \citealt{2021A&A...648A..99T} for more details]{2020A&A...644A..15M} and \textsc{QSFit} \citep{2017MNRAS.472.4051C}. Specifically, the SDSS data and the two LR-B spectra were fitted over the wavelength range 4500--8000 \AA, while the VHR-R one over the range 6250--6800 \AA. Being the results obtained with the two different fitting algorithms fully consistent, we adopt the best-fit results from the fitting code by \citet{2020A&A...644A..15M}, which allows to fix emission-line components of different lines to the same kinematics.

The fitting code by \citet{2020A&A...644A..15M} consists of two main steps. We first modelled the full spectra with \textsc{pPXF} \citep{2004PASP..116..138C, 2017MNRAS.466..798C}, building dedicated templates for the various spectral components: namely, the stellar continuum, the AGN continuum, and the emission lines from the AGN BLR and NLR. After that, we subtracted from the data the overall continuum (i.e. stellar and AGN continuum) and the BLR emission, and performed a more accurate modelling of the narrow emission lines only, aiming to check whether the narrow-line emission is subject to variability as well. Details on the two steps of the spectral modelling are provided below.

\begin{figure}
\centering\includegraphics[width=\hsize]{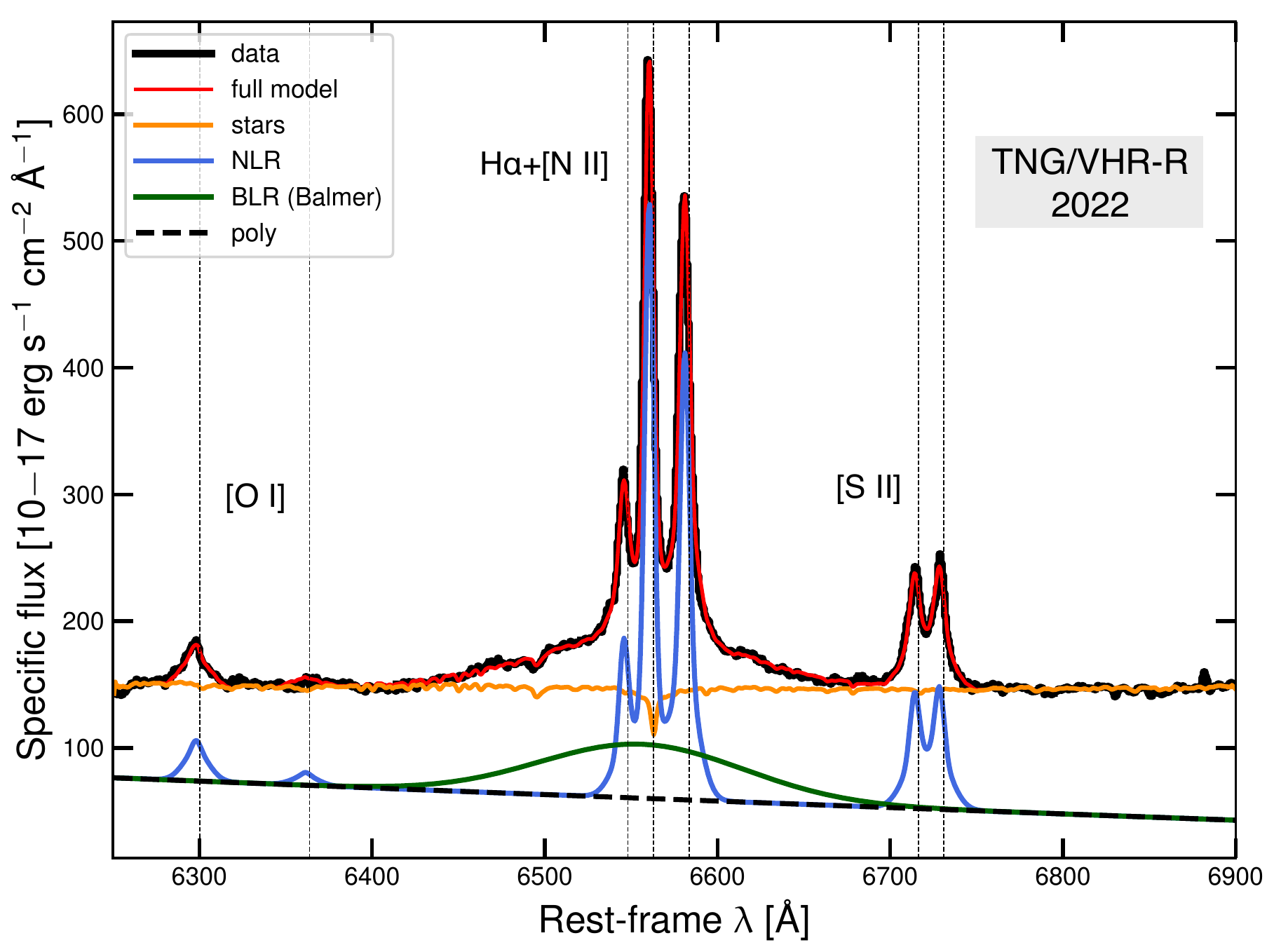} 
    \caption{Best-fit modelling of the TNG/VHR-R spectrum of NGC 4156, as a representative example of our first fit with \textsc{pPXF}. Data are shown as solid black line along with the full best-fit model (red) and the separate models for each spectral component (different colours, see the legend). A 1st-degree polynomial or a broken power law (dashed black line) is used to reproduce the AGN contribution to the continuum emission.}
    \label{fig:full_fit}
\end{figure}

Figure \ref{fig:full_fit} shows the observed VHR-R 2022 spectrum (black) as a representative example of the first phase of our spectral modelling. The total best-fit model and the single templates for each spectral component are superimposed in red and different colours, respectively. The stellar continuum was modelled using the MILES extended stellar population models \citep{2016A&A...589A..73R}, whereas the semi-analytic templates of \citet{2010ApJS..189...15K} were used to reproduce the broad [\ion{Fe}{ii}] emission lines originating in the AGN BLR. A prominent [\ion{Fe}{ii}] emission component is required to properly model the steep increase of the continuum emission in the bluest part of the LR-B 2019 spectrum (Fig. \ref{fig:full_fit_LRB19}), but it is absent in the other spectra, except for a paltry contribution to the total emission around 5300\AA\ in the LR-B 2022 spectrum (Fig. \ref{fig:full_fit_LRB22}).

In general, \textsc{pPXF} gives the possibility of including an additive nth-degree polynomial to either adjust the continuum shape to the observed data, or account for any additional continuum component. Whereas a 1st-degree polynomial was sufficient to properly reproduce the faint AGN continuum in the SDSS and VHR-R spectra, it was not appropriate for the steeply rising continuum in the bluest part of the 2019 spectrum. Therefore, we created a set of broken power law templates with independently-variable slopes ($F_\lambda\propto \lambda^\alpha$, with $-3.5<\alpha_1,\alpha_2<0$). To reduce the number of templates, we fixed the break point of all templates to 5800 \AA, as resulting from the 2019 best-fit obtained with \textsc{QSFit}. In the LR-B 2022 data instead, the AGN continuum cannot be well constrained, being overwhelmed by the almost flat stellar continuum over the entire wavelength range 4500--8000 \AA\ (Fig. \ref{fig:full_fit_LRB22}). Nonetheless, we expect an AGN continuum component to be present as well, since the LR-B 2022 continuum remains at a higher flux level with respect to the SDSS 2004 one (see Fig. \ref{fig:comparison}), which cannot be due to a variation in the stellar component. The comparison between the 2022 LR-B and VHR-R best-fit results for the AGN continuum further points out the degeneracy of this template in the 2022 data, especially its strong dependence on the fitted wavelength range (cfr. Figs. \ref{fig:full_fit} and \ref{fig:full_fit_LRB22}).

The main emission lines detected in the covered wavelength range are the Balmer hydrogen lines \ha\ and \hb, and the forbidden lines [\ion{O}{iii}]$\lambda\lambda$4959,5007,  [\ion{O}{i}]$\lambda\lambda$6300,64, [\ion{N}{ii}]$\lambda\lambda$6549,83, and [\ion{S}{ii}]$\lambda\lambda$6716,31. All the emission lines were modelled by \textsc{pPXF} using multiple Gaussian components, with the [\ion{O}{iii}], [\ion{O}{i}] and [\ion{N}{ii}] fitted as doublets with a fixed flux ratio of 3 between the two components. Hereafter we refer to the brighter doublet component (i.e. [\ion{O}{iii}]$\lambda$5007, [\ion{O}{i}]$\lambda$6300 and [\ion{N}{ii}]$\lambda$6583) as [\ion{O}{iii}], [\ion{O}{i}] and [\ion{N}{ii}], respectively.
We found that two narrow ($\sigma$ of a few hundreds km s$^{-1}$) Gaussian components are needed to properly reproduce the NLR contribution to each emission line; whereas an additional broad ($\sigma>2000$ km s$^{-1}$) Gaussian component was included to account for any possible BLR contribution to the \ha\ and \hb\ line profiles (not present at all in the SDSS spectrum, see Fig. \ref{fig:full_fit_SDSS}). Each set of (broad or narrow) Gaussian components was additionally constrained to have the same kinematics (i.e. same $v$ and $\sigma$).

We point out a significant degeneracy amongst the faint broad \hb\ and the other spectral components that significantly contribute over the \hb\ wavelength range. These are mainly the AGN continuum and [\ion{Fe}{ii}] emission in the 2019 spectrum; and the stellar continuum in the LR-B 2022 data. Constraining the broad \hb\ to the same kinematics of the brighter, broad \ha\ allowed us to better disentangle the broad \hb\ component. While the resulting flux of the 2022 broad \hb\ appears to be reasonable compared to that of the 2022 broad \ha, the 2019 broad \hb\ flux is likely an overestimate, given the unreliably small value of the Balmer decrement (BD; [H$\alpha$/H$\beta$]$^{\rm B,2019}\sim 1.7$) compared to that computed for the narrow components ([H$\alpha$/H$\beta$]$^{\rm N,2019}=4.5\pm1.4$) and the theoretical one ($\sim$ 2.87). In the following, all the luminosity/flux values are corrected for intrinsic dust extinction \citep{2000ApJ...533..682C}: for the 2019 estimates we adopted the 2019 NLR BD, whereas for the 2022 ones the 2022 BLR BD ([H$\alpha$/H$\beta$]$^{\rm B,2022}=4\pm2$). A 50\% uncertainty has been considered on the BLR BD value to account for the systematic errors due to the uncertainty on the BLR components, strongly depending on the continuum best-fit. As a consequence, the narrow-line measurements are expected to be affected as well although to a lesser extent ($\sim$30\% uncertainty assumed on the NLR BD).

To check whether any variability affects the narrow-line emission as well, we rely on the refined modelling of the data after subtracting the overall continuum and BLR ([\ion{Fe}{ii}] and broad Balmer components) emission. Similarly to the previous step, we fitted a multiple Gaussian model to each emission line, constraining the components of each Gaussian set to the same kinematics. Two Gaussian components were again required to properly reproduce the narrow-line emission profiles, except for the LR-B 2022 spectrum, which required a third component to optimally reproduce simultaneously all the emission lines, especially the [\ion{S}{ii}] and [\ion{O}{iii}] doublets (see Appendix \ref{apx:spfits}).

All the results from the spectral fitting are described in Appendix \ref{apx:spfits}, with the main parameters listed in Tables \ref{tab:sp_balmer_broad}, \ref{tab:sp_balmer_nar}, \ref{tab:sp_forbidden_oiii} and \ref{tab:sp_forbidden} along with their statistical errors. We caution that such errors are likely underestimating the true uncertainty, especially on the broad \ha\ and \hb\ values (see Appendix \ref{apx:spfits}).

\section{Results}
\label{sec:results}

The key result of this work is the intriguing discovery of NGC 4156 as CLAGN, as pointed out by the appearance of BLR \ha\ and \hb\ components in 2019, which suggest a global transition from a type 2 towards a type 1 since 2004. In addition, the appearance (in 2019) and subsequent disappearance (in 2022) of the blue rising continuum hints at a further evolution backwards to a type 2, overall making this galaxy extremely interesting to study.

On the contrary, no significant variation of the narrow emission-line flux ratios has occurred from 2004 to 2022. All different-epoch narrow flux ratios indeed lie in the same region of the Baldwin, Phillips \& Terlevic (BPT; \citealt{1981PASP...93....5B}) diagrams (see Figure \ref{fig:bpt} and related text).


Thanks to the appearance of BLR emission, for the first time, we can use optical data to compute fundamental AGN properties such as black hole mass ($M_{\rm BH}$), bolometric luminosity ($L_{\rm bol}$), Eddington ratio ($\lambda_{\rm Edd}=L_{\rm bol}/L_{\rm Edd}$), and characteristic radius of the BLR ($R_{\rm BLR}$).

We derive the single-epoch estimate of $M_{\rm BH}$ from the VHR-R 2022 data, whose higher resolution enables more accurate measurements of the broad \ha\ component. Following the prescription by \citet{2005ApJ...630..122G}, we estimate $M_{\rm BH}$ to be equal to $(1.4\pm0.1)\times10^8$ \(M_\odot\) (i.e. $\log(M_{\rm BH}/\text{\(M_\odot\)})\sim8.1$) from the VHR-R $\sigma^{\rm B}_{\rm H\alpha}$ ($\simeq2690$ km s$^{-1}$) and broad \ha\ luminosity ($L^{\rm B}_{\rm H\alpha}\sim1.9\times10^{41}$ erg s$^{-1}$). Such a value is remarkably similar to the one predicted by the $M_{\rm BH}-\sigma_*$ relation based on the SDSS data.

For the 2019 and 2022-epoch estimates of $L_{\rm bol}$, we apply an optical bolometric correction at 4400\AA\ of 5.0 \citep[][]{2020A&A...636A..73D} to the AGN continuum luminosity ($\lambda L_{4400}$). Given the difficulty in constraining the AGN continuum directly from the spectral modelling (discussed in Sect. \ref{sec:spanalysis} and Appendix \ref{apx:spfits}), we compute $\lambda L_{4400}$ by making the following assumption. Since we do not expect that the stellar continuum has varied since 2004, we subtracted the best-fit stellar continuum obtained from the 2004 spectrum to the LRB 2019 and 2022 ones. We then consider any additional emission at 4400 \AA\ as the AGN continuum. From the
values of $\lambda L_{4400}$ ($\lambda L^{\rm 2019}_{4400}\simeq4.8\times10^{43}$ erg s$^{-1}$ and $\lambda L^{\rm 2022}_{4400}\simeq4.1\times10^{42}$ erg s$^{-1}$), we obtain the bolometric luminosity values of $L^{\rm 2019}_{\rm bol}=(2.4\pm1.2)\times10^{44}$ erg s$^{-1}$ and $L^{\rm 2022}_{\rm bol}=(2.1\pm1.1)\times10^{43}$ erg s$^{-1}$, where 50\% errors are driven by the uncertainty on the bolometric correction.
From the inferred values of $M_{\rm BH}$ and $L_{\rm bol}$, we compute the Eddington ratio corresponding to the epochs 2019 and 2022, $\lambda^{\rm 2019}_{\rm Edd}\simeq0.015$ and $\lambda^{\rm 2022}_{\rm Edd}\simeq0.001$, respectively.

\citet{elitzur2014} proposed that the AGN broad-line emission follows an evolutionary sequence from type 2 to 1 (and backwards), through intermediate states, as the SMBH accretion rate increases (decreases). Such spectral evolution is described by the parameter $L_{\rm bol}/M^{2/3}_{\rm BH}$ ($L_{\rm bol}$ and $M_{\rm BH}$ in units of erg s$^{-1}$ and \(M_\odot\), respectively), where values larger than 35 are typical of BLR emission. From a qualitative comparison with Fig. 1 in \citet{elitzur2014}, we expect NGC 4156 to be compatible with a type 1.2/1.5 in 2019 ($\log(L^{\rm 2019}_{\rm bol}/M^{2/3}_{\rm BH})\simeq39.0$), whereas more with a type 1.8 in 2022 ($\log(L^{\rm 2022}_{\rm bol}/M^{2/3}_{\rm BH})\simeq37.9$), given the factor of $\sim$10 of difference between our 2019 and 2022 $L_{\rm bol}$ estimates. The rough type 1.8 classification for the 2022 epoch is also supported by the 2022 value of $L^{\rm B}_{\rm H\alpha}\sim10^{41}$ erg s$^{-1}$ \citep{sl2012}. This suggests that in 2019 and 2022 we might have observed NGC 4156 in two distinct intermediate phases of its round-trip evolution along the type sequence.

We finally estimate $R_{\rm BLR}$ from the AGN continuum luminosity at 5100 \AA\ ($\lambda L_{5100}$) via the \hb\ reverberation time lag calibrated by \citet{bentz2013}.
By taking our best constrained value of $\lambda L_{5100}$, that is from the 2019 best-fit model of AGN continuum ($\lambda L_{5100}\simeq1.3\times10^{43}$ erg s$^{-1}$), we find $R_{\rm BLR}\sim11$ light days.
Since the NLR likely extends up to a few hundreds of parsecs \citep[see Eqs 1, 2 and 3][]{bennert2002}, the narrow lines are expected to vary on a time scale much longer than the time frame spanned by these observations (2004--2022), consistently with no significant change observed in the narrow-line emission of NGC 4156 since 2004. 

\section{Discussion and conclusions}
\label{sec:discussion}
In this Letter, we report on the serendipitous discovery of NGC 4156 as a CLAGN, undergoing a double type transition: first, from a type 2 towards a type 1 (1.5 in 2019), and now getting back to a type 2 (1.8 in 2022). Optical TNG data taken in 2019 have indeed revealed the appearance of a BLR component in the \ha\ and \hb\ line profiles, together with a rising blue continuum. The BLR components have been confirmed by our 2022 follow-up observations, whereas the rising continuum has disappeared. In both 2019 and 2022, the broad \ha\ is bright and well detected, whilst the broad \hb\ emission line is barely detected and harder to identify, especially in 2022 (see Sect. \ref{sec:spanalysis} and Appendix \ref{apx:spfits}). This suggests that in 2019 and 2022 we have caught NGC 4156 in two intermediate states \citep{o1981} of its 2-to-1 type round-trip evolution.

The spectral evolution and type transitions observed in the optical over the last twenty years might be the result of variable dust absorption, shielding the central AGN along the line of sight, or changes in the accretion rate. To test the variable absorption scenario, we compute the crossing time ($t_{\rm cross}$) of an intervening screen orbiting the BLR. Assuming a Keplerian orbit of radius $R_{\rm cloud}=3R_{\rm BLR}$ \citep{lamassa2015} and our \ha-based \mbh\ estimate, we obtain $t_{\rm cross}\sim4$ years. Since NGC 4156 has not yet fully transitioned back to a type 2 in a time of 3 years (i.e. 2019-2022), the shielding of the nucleus seems to be a viable scenario. Even though we have approximate estimates of the NLR BDs in all epochs, such values are not necessarily representative of the BLR extinction, which is unknown in 2004 and 2019, and might also affect the continuum. We can only estimate the 2022 BLR BD, considering a $\sim$50\% systematic uncertainty (discussed in Sect. \ref{sec:spanalysis}).

On the other hand, the inferred 2019 and 2022 values of $L_{\rm bol}$ and $\lambda_{\rm Edd}$ suggest a decreased accretion activity in NGC 4156, which might explain the recent flattening of the continuum emission. Moreover, a systematic analysis of a large number of intermediate-type AGN with $L_{\rm bol} < 10^{44}$ erg s$^{-1}$ \citep[e.g.][]{sl2012} has demonstrated that these systems should be classified as low-luminosity unobscured AGN rather than as obscured type 2 objects and, consequently, any observed change possibly ascribed to a variable accretion rate.
In spite of the unlikely nature of NGC 4156 as a fully unobscured AGN (BD $> 3.5$ in all epochs), its 2019/2022 emission is consistent with that of optically-selected type 1 AGN in the $L_{\text{[\ion{O}{iii}]}}-L^{\rm B}_{\rm H\alpha}$ plane \citep[Figure 8 in][]{sl2012}. Except for the 2019 bright phase, the low-luminosity AGN nature of NGC 4156 is also supported by the low contrast between the AGN and the host galaxy, with the latter overwhelming the AGN continuum in 2004 and 2022 data.

Further insights into the observed variability of this source may arise at X-ray energies. At the time of our 2022 TNG follow-up, there were no archival X-ray pointed observations of NGC 4156, which, however, falls in the field of NGC 4151 in 9 different \xmm observations (from 2000 to 2012). A preliminary analysis of these data reveals some degree of X-ray variability of NGC 4156.
Interestingly, new \xmm data (June 2022, PI: Lusso) show no clear signs of absorption. Therefore, if absorption is the main responsible for the observed change in the optical, such obscuring material is not attenuating the flux at higher energies. We defer a more detailed analysis of the X-ray properties of NGC 4156 to a dedicated publication. 


In conclusion, the discovery of NGC 4156 as a CLAGN makes this galaxy worthy of further investigations. An optical and X-ray monitoring of this galaxy would allow us to constrain the duty cycle of the nuclear activity and to shed light on the mechanism driving the type transition in NGC 4156, clarifying the role played by variable dust absorption. Such a monitoring would also reveal whether (and possibly when) the galaxy will revert to a type 2. Polarization monitoring campaigns might also bring crucial information on the geometry and composition of the nuclear reprocessing regions as well as on the AGN inclination \citep[e.g.][]{1989ApJ...340..190G,h2019,mh2020}.



\begin{acknowledgements}
We wish to thank 
the referee and 
B. M. Peterson for insightful comments.
Based on observations made with the Italian Telescopio Nazionale Galileo (TNG) operated by the Fundaci\'on Galileo Galilei (FGG) of the Istituto Nazionale di Astrofisica (INAF) at the Observatorio del Roque de los Muchachos (La Palma, Canary Islands, Spain).
We thank the TNG Director E.\ Poretti for Director Discretionary Time allowing us to perform the 2022 follow-up observations (PI: E.\ Lusso), for which we also thank M.\ Cecconi as support astronomer. 
The participation of the students to the 2019 TNG observing campaign was financially supported by INAF-Osservatorio Astrofisico di Arcetri, through the Magini's fund, and by the Dipartimento di Fisica e Astronomia, Università di Firenze. We thank again the TNG Director and the whole TNG staff for making such an experience possible. 

\end{acknowledgements}

%
   \bibliographystyle{aa} 
   \bibliography{biblio} 
%
\begin{appendix} 
\section{The students' observing campaign}
\label{apx:obscampaign}
The observation of NGC 4156 carried out at the TNG on April 28$^{\rm th}$, 2019 was part of a multi-year campaign planned and performed by undergraduate students of the Physics and Astronomy Department of the University of Florence (Italy), within a course of introductory observational astrophysics \citep{Head2022}. To date, this programme has  targeted 89 nearby ($z<0.2$), bright active and non-active spiral galaxies to perform both imaging and low-resolution spectroscopy with different facilities. Since 2010, 74 galaxies have been observed at the 1.5m Cassini telescope in Loiano (operated by the INAF Bologna, Italy), 2 at the 1.8m Copernico telescope in Asiago-Ekar (INAF Padova, Italy), and 13 at the 3.6m TNG telescope.
The main goals of the observations were the measurements of the galaxies' apparent diameter (from imaging) and redshift (from spectroscopy), to estimate the Hubble parameter $H_0$. Additional goals were the estimates of the star formation rate and of $M_{\rm BH}$, in case of active galaxies. 

As briefly mentioned in Sec.\ \ref{sec:obs}, the primary target of the 2019 spectroscopic observation was not NGC 4156 itself, but the brighter, nearby galaxy NGC 4151. Given the mutual proximity of the two galaxies and the large enough field of view of Dolores, we acquired the spectra of both galaxies by placing the slit along the direction joining the two nuclei, thus increasing the number of observed sources. However, while the NGC 4151 data were immediately reduced (showing no unexpected features), the data of NGC 4156 were not. Therefore, the surprising change in the spectral features of NGC 4156 has remained undiscovered until the first data reduction eventually performed in 2022.    

\section{Spectral fits results}
\label{apx:spfits}

\begin{table*}
	\centering
	\caption{Observed flux, EW and $\sigma$ of the BLR component of the \ha\ and \hb\ lines, as resulting from our 1st-phase modelling (Sect. \ref{sec:spanalysis}). The physical quantities are in units of 10$^{-15}$ erg s$^{-1}$ cm$^{-2}$, \AA~ and km s$^{-1}$, respectively. The 2022/LR-B$^{\dagger}$ row reports the parameters for the broad \ha\ obtained from the re-modelling of the LR-B 2022 spectrum over the VHR-R wavelength range (i.e. $6250-6900$ \AA).} 
	\label{tab:sp_balmer_broad}
	\centering
	\begin{tabular}{lccccc} 
		\hline
		Year & $F^{\rm B}_{\rm H\beta}$ & EW$^{\rm B}_{\rm H\beta}$ & $F^{\rm B}_{\rm H\alpha}$ & EW$^{\rm B}_{\rm H\alpha}$ & $\sigma^{\rm B}_{\rm H\alpha, \beta}$\\
		\hline
		2004/SDSS & - & - & - & - & - \\
		2019/LR-B & 37.1$\pm$0.2 & 22.2$\pm$0.4 & 62.6$\pm$0.2 & 42.0$\pm$0.7 & 3690$\pm$16\\
		2022/LR-B & 9.2$\pm$0.3 & 6.5$\pm$0.6 & 37.1$\pm$0.4 & 24$\pm$2 & 2110$\pm$30\\
		2022/LR-B$^{\dagger}$ & - & - & 56.8$\pm$0.4 & 38$\pm$4  & 2460$\pm$30\\
		2022/VHR-R & - & - & 64.6$\pm$0.2 & 42.9$\pm$0.3 & 2694$\pm$17\\
		\hline
	\end{tabular}
\end{table*}

\begin{table*}
	\centering
	\caption{Same as Table \ref{tab:sp_balmer_broad} for the NLR component of the \ha\ and \hb\ lines.}

	\label{tab:sp_balmer_nar}
	\centering
	\begin{tabular}{lcccccc}
		\hline
		Year & $F^{\rm N}_{\rm H\beta}$ & EW$^{\rm N}_{\rm H\beta}$ & $\sigma^{\rm N}_{\rm H\beta}$ & $F^{\rm N}_{\rm H\alpha}$ & EW$^{\rm N}_{\rm H\alpha}$ & $\sigma^{\rm N}_{\rm H\alpha}$\\
		\hline
		2004/SDSS & 6.2$\pm$0.3 & 5.1$\pm$0.2 & 220$\pm$10 & 28.5$\pm$0.6 & 21.4$\pm$0.3 & 230$\pm$10\\
		2019/LR-B & 8$\pm$3 & 4.8$\pm$0.1 & 310$\pm$40 & 36$\pm$11 & 24.3$\pm$0.3 & 300$\pm$40\\
		2022/LR-B & 8.9$\pm$1.6 & 6.3$\pm$0.1 & 360$\pm$90 & 32$\pm$5 & 21$\pm$6 & 310$\pm$90\\
		2022/VHR-R & - & - & - & 37.4$\pm$0.6 & 24.8$\pm$0.1 & 220$\pm$10\\
		\hline
	\end{tabular}
\end{table*}

\begin{table*}
	\centering
	\caption{Same as Table \ref{tab:sp_balmer_broad} for the [\ion{O}{iii}] emission line.}

	\label{tab:sp_forbidden_oiii}
	\begin{tabular}{lccc} 
		\hline
		Year & $F_{\rm [\text{\ion{O}{iii}}]}$ & EW$_{\rm[ \text{\ion{O}{iii}}]}$ & $\sigma_{\rm[ \text{\ion{O}{iii}}]}$\\
		\hline
		2004/SDSS & 7.6$\pm$0.3 & 6.2$\pm$0.4 & 290$\pm$10\\
		2019/LR-B & 10.6$\pm$3 & 6.3$\pm$0.1 & 310$\pm$40\\
		2022/LR-B & 12$\pm$2 & 8.4$\pm$0.1 & 440$\pm$90\\
		2022/VHR-R & - & - & -\\
		\hline
	\end{tabular}
\end{table*}

\begin{table*}
	\centering
	\caption{Same as Table \ref{tab:sp_balmer_broad} for the main forbidden emission lines in the wavelength range $6400-6800$ \AA.}
	\label{tab:sp_forbidden}
	\begin{center}
	\begin{tabular}{lccccccccc} 
		\hline
		Year & $F_{\rm [\text{\ion{N}{ii}}]}$ & EW$_{\rm[ \text{\ion{N}{ii}}]}$ & $\sigma_{\rm[ \text{\ion{N}{ii}}]}$ & $F_{\rm [\text{\ion{S}{ii}}]6716}$ & EW$_{\rm[ \text{\ion{S}{ii}}]6716}$ & $\sigma_{\rm[ \text{\ion{S}{ii}}]6716}$ & $F_{\rm [\text{\ion{S}{ii}}]6731}$ & EW$_{\rm[ \text{\ion{S}{ii}}]6731}$ & $\sigma_{\rm[ \text{\ion{S}{ii}}]6731}$\\
		\hline
		2004/SDSS & 28.8$\pm$0.5 & 22.0$\pm$0.3 & 280$\pm$10 & 9.1$\pm$0.4 & 7.0$\pm$0.3 & 270$\pm$10 & 9.4$\pm$0.4 & 7.2$\pm$0.3 & 290$\pm$10\\
		2019/LR-B & 30$\pm$10 & 21.0$\pm$0.2 & 300$\pm$40 & 11$\pm$5 & 7.8$\pm$0.1 & 290$\pm$40 & 11$\pm$4 & 7.9$\pm$0.1 & 300$\pm$40\\
		2022/LR-B & 29$\pm$5 & 18.7$\pm$0.6 & 360$\pm$90 & 76$\pm$2 & 5.1$\pm$0.4 & 330$\pm$90 & 8$\pm$4 & 5.4$\pm$0.7 & 320$\pm$90\\
		2022/VHR-R & 34.9$\pm$0.4 & 24.2$\pm$0.1 & 270$\pm$10 & 9.7$\pm$0.4 & 6.7$\pm$0.1 & 290$\pm$10 & 10.7$\pm$0.4 & 7.4$\pm$0.1 & 290$\pm$10\\
		\hline
	\end{tabular}
	\end{center}
\end{table*}

The main results from the spectral analysis (i.e. observed flux, equivalent width and velocity dispersion of the emission lines) of the four optical spectra of NGC 4156 (i.e. the archival SDSS 2004, and our three TNG spectra) are summarised in Tables \ref{tab:sp_balmer_broad}, \ref{tab:sp_balmer_nar}, \ref{tab:sp_forbidden_oiii} and \ref{tab:sp_forbidden}. Tables \ref{tab:sp_balmer_broad} and \ref{tab:sp_balmer_nar} refer to the hydrogen Balmer lines \ha\ and \hb, with the former showing the measurements for the BLR component, while the latter for the NLR component. Instead, Tables \ref{tab:sp_forbidden_oiii} and \ref{tab:sp_forbidden} show the results for the forbidden emission lines. The uncertainty reported for the estimated quantities are the statistical errors resulting from the fit, which are likely to underestimate the real uncertainty. This is true especially for the broad \ha\ and \hb\ measurements, which strongly depend on the continuum. Since the narrow-line emission has been modelled with multiple Gaussian components, the narrow $\sigma$ values have been computed as moment-2 of the overall narrow line profile. As mentioned in Sect. \ref{sec:spanalysis}, the LR-B 2022 data required 3 Gaussian components to properly reproduce simultaneously the narrow profiles of all emission lines. In doing so, we correctly modelled both a faint blue wing in the [\ion{O}{iii}] profile and the peak of the blended [\ion{S}{ii}] doublet, whose two components are instead well resolved in the LR-B 2019 due to the narrower aperture of the slit (i.e. 1$^{\prime\prime}$ against 1.5$^{\prime\prime}$ in 2022), hence higher spectral resolution.

We notice that the broad \ha\ flux and EW values obtained from the analysis of the VHR-R 2022 spectra are larger by a factor of $\sim$1.7 than the corresponding LR-B 2022 values. Such a difference is due to the different resolution and fitted wavelength range. The higher-resolution VHR-R data indeed allow us to detect and model even the shallower wings of the broad \ha, and also to better decompose the \ha$+$[\ion{N}{ii}] line complex. In fact, by re-modelling the LR-B 2022 spectrum only over the VHR-R wavelength range (i.e. $6250-6900$ \AA; see lower right panel of Fig. \ref{fig:full_fit_LRB22}), we obtained a best-fit globally similar to the VHR-R one: the level of the AGN continuum has changed, being now reproduced by a similar 1st-degree polynomial, and the broad \ha\ component has become consistent with its corresponding VHR-R counterpart. The best-fit results for the broad \ha\ from this re-modelling are referred to as 2022/LR-B$^{\dagger}$ in Table \ref{tab:sp_balmer_broad}. This simple check demonstrates how sensitive the broad components are to the level of the total continuum, which in turn strongly depends on the fitted wavelength range (compare the right panels of Fig. \ref{fig:full_fit_LRB22}).

By comparing the results from different epochs, the narrow emission is not seen to significantly vary, whereas we notice a 2022 broad \hb\ flux smaller by a factor of $\sim$4 than in 2019 (consequently, the EW as well). This apparent dimming is unlikely real, but rather a consequence of the degeneracy between the broad \hb\ and the spectral components (such as [\ion{Fe}{ii}]) giving a substantial contribution in the wavelength range around the \hb\ line. Moreover, this part of the spectrum (including the \hb\ line) is expected to be more extincted compared to the spectral range of the \ha\ line, which is also intrinsically brighter than \hb. Therefore, it is hard to disentangle and properly identify the real broad \hb\ component. In the specific case of the LR-B 2019 data, it is likely that we have overestimated the flux of the broad \hb\ component, as suggested by the unreliably small value of [H$\alpha$/H$\beta$]$^{\rm B,2019}\sim1.7$ compared to the NLR BD ([H$\alpha$/H$\beta$]$^{\rm N,2019}=4.5\pm1.4$) and the theoretical ($\sim2.87$) values. Similarly, we caution also on the 2022 broad \hb\ measurements, with a [H$\alpha$/H$\beta$]$^{\rm B,2022}=4.0\pm2.0$, anyway pointing to a significant level of extinction. For all these reasons, we eventually used the VHR-R \ha\ measurements to infer the main AGN properties (such as $M_{\rm BH}$; see Sect. \ref{sec:results}).

\begin{figure*}
\centering\includegraphics[width=0.95\hsize]{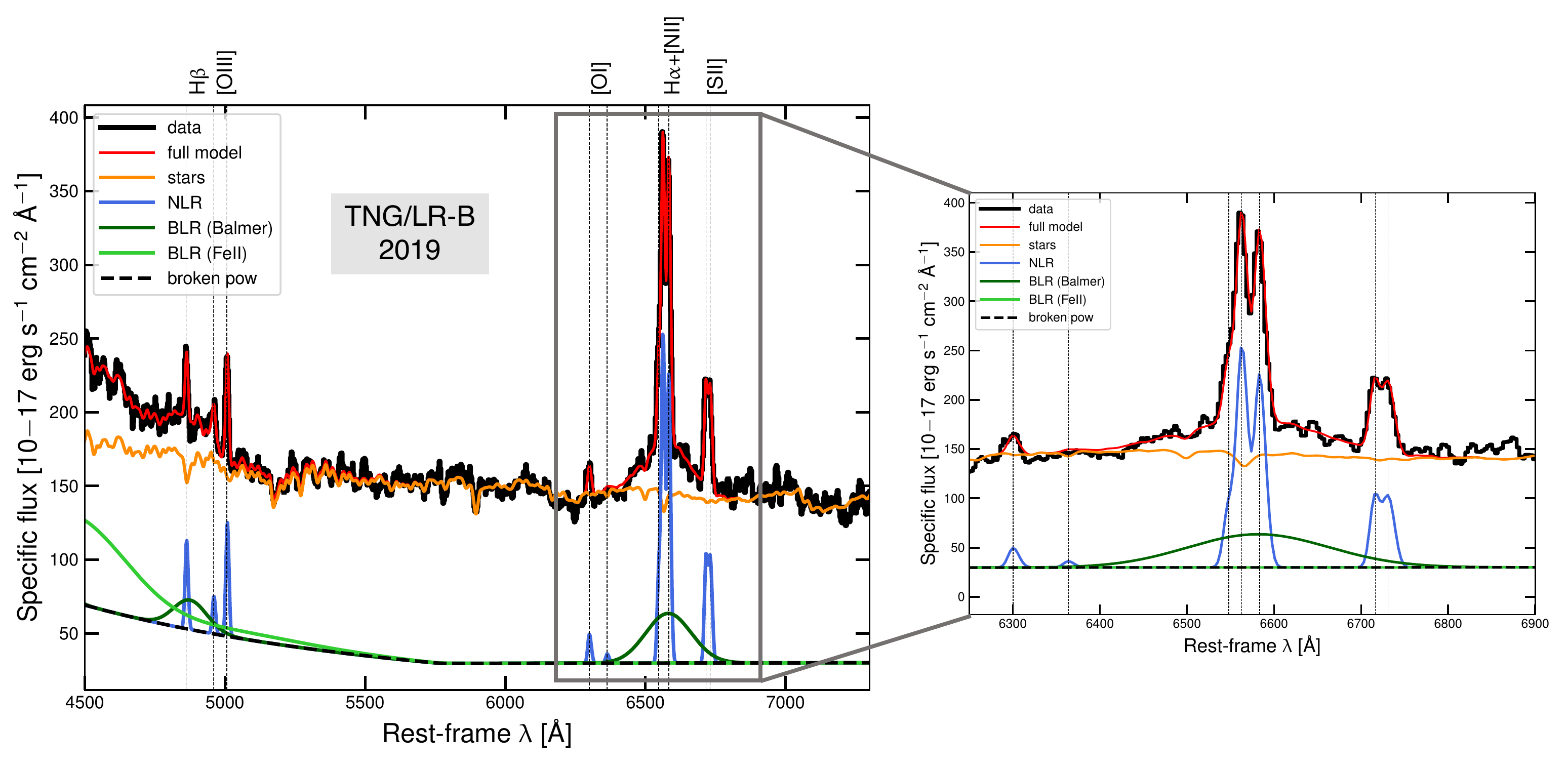} 
    \caption{\textit{Left.} Best-fit modelling resulting from our 1st-phase fit of the TNG/LR-B 2019 spectrum. Data are shown as a solid black line along with the full best-fit model in red, and separate models for the various spectral components in different colours (see the legend). A broken power law (dashed black line) has been used to approximate the AGN continuum. \textit{Right.} Zoom-in over the spectral window $6250-6900$ \AA\ around the \ha\ line.}
    \label{fig:full_fit_LRB19}
\end{figure*}

\begin{figure*}
\centering\includegraphics[width=0.95\hsize]{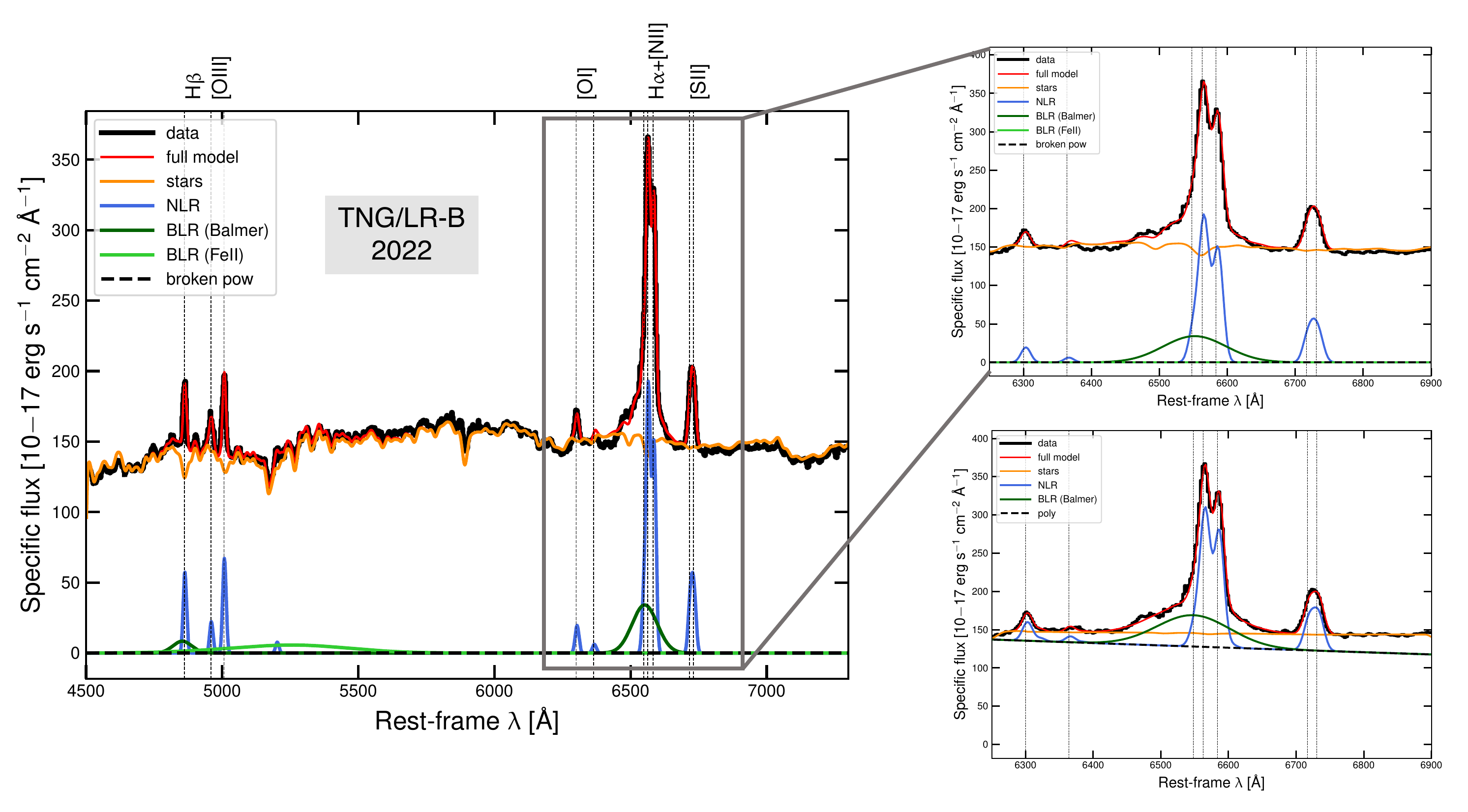} 
    \caption{\textit{Left.} Same as Fig. \ref{fig:full_fit_LRB19} for the TNG/LR-B 2022 spectrum. The AGN continuum is totally unconstrained and overwhelmed by the host galaxy emission. \textit{Upper right.} Zoom-in over the spectral window $6250-6900$ \AA\ around the \ha\ line. \textit{Lower right.} Best-fit results from the re-modelling over the narrower wavelength range (i.e. $6250-6900$) of the VHR-R data. The comparison between the two right panels shows how sensitive the broad components are to the continuum level, which in turns strongly depends on the spectral range selected for the fit.}
    \label{fig:full_fit_LRB22}
\end{figure*}

\begin{figure*}
\centering\includegraphics[width=0.95\hsize]{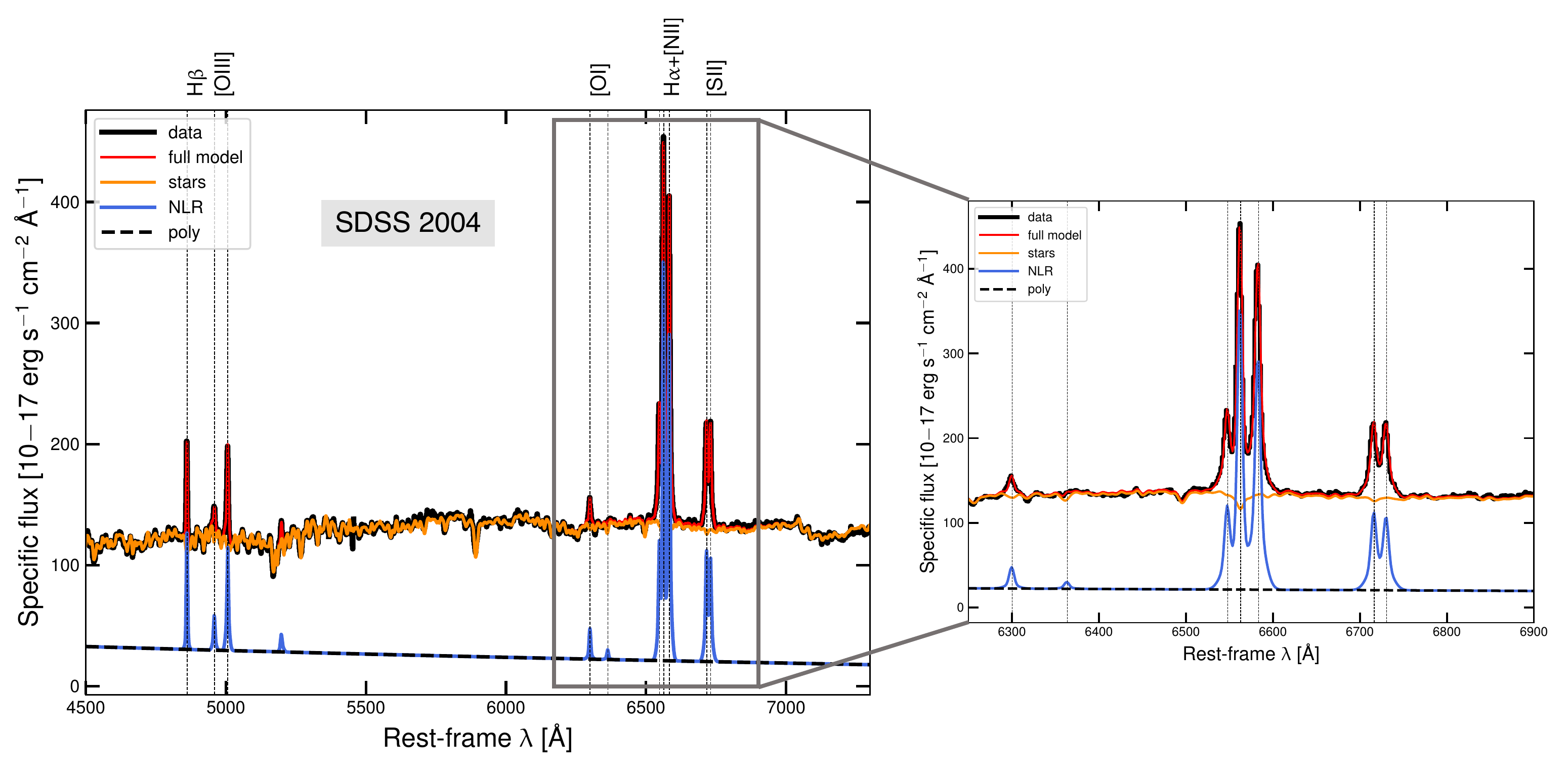} 
    \caption{\textit{Left.} Same as Fig. \ref{fig:full_fit_LRB19} for the SDSS 2004 spectrum. Data are shown as solid black line along with the full best-fit model in red, and separate models for the various spectral components in different colours (see the legend). A slight 1st-degree polynomial (dashed black line) is fitted to the weak AGN continuum, while no BLR components are detected. \textit{Right.} Zoom-in over the spectral window $6250-6900$ \AA\ around the \ha\ line, clearly showing the total absence of a broad \ha\ component.}
    \label{fig:full_fit_SDSS}
\end{figure*}

\section{BPT diagram}
\label{BPT diagram}
\begin{figure*}
\centering\includegraphics[width=0.9\hsize]{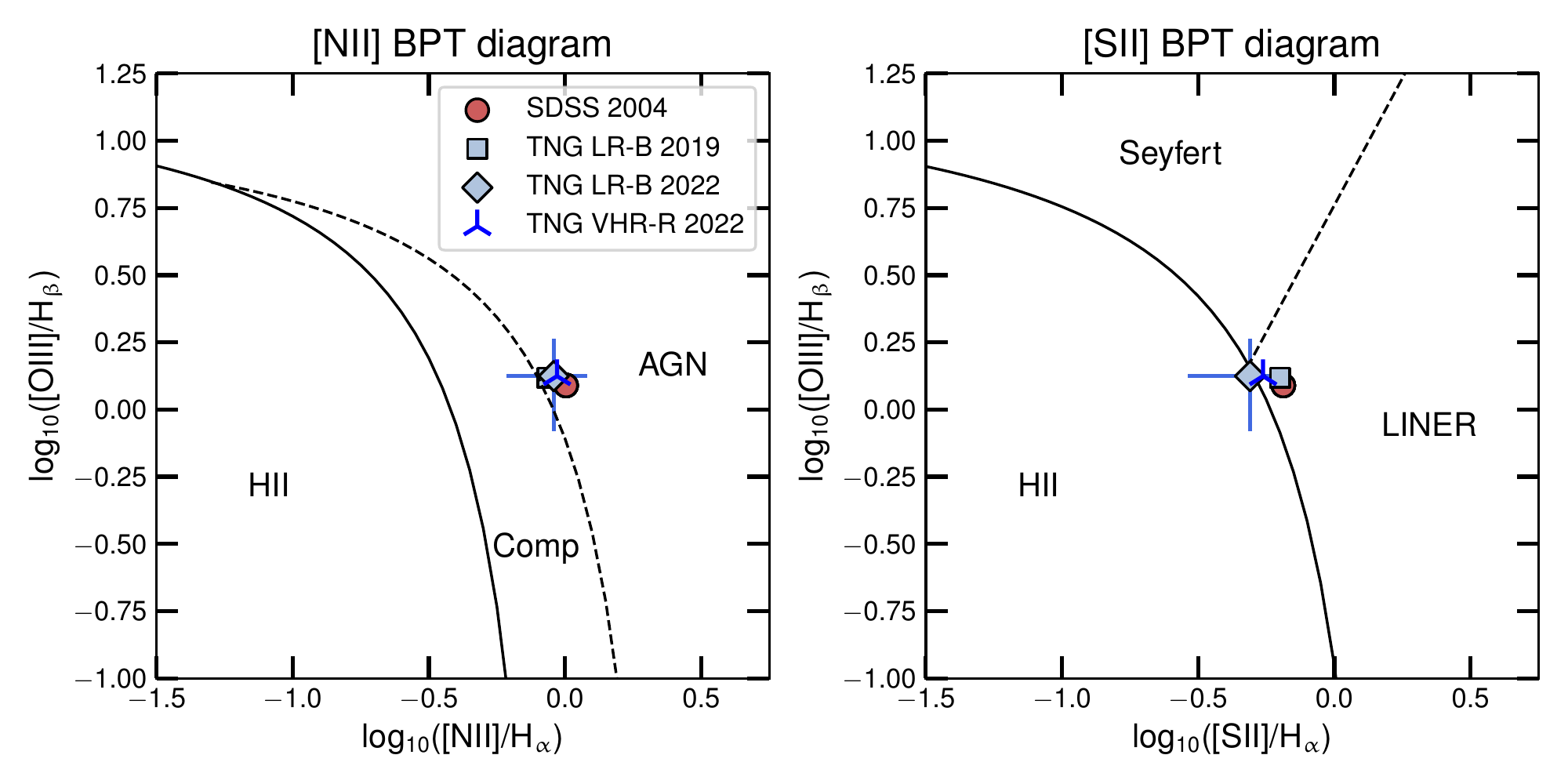} 
    \caption{[\ion{N}{ii}]- and [\ion{S}{ii}]-BPT diagrams with line-flux ratio measurements for the 4 spectra of NGC 4156: archival SDSS 2004 (circle) and our TNG LR-B 2019 (square), LR-B 2022 (diamond) and VHR-R 2022 (blue marker) spectra. The VHR-R 2022 points have been obtained by combining the measurements of \textit{x}-axis and \textit{y}-axis ratios from the VHR-R and LR-B 2022 spectra, respectively. Solid and dashed black lines, separating between different ionisation regimes, are taken from \citet{2001ApJ...556..121K, 2006MNRAS.372..961K} and \citet{2003MNRAS.346.1055K}. For clarity, we plot only the error bars relative to the TNG/LR-B 2022 measurements, comparable to the TNG/LR-B 2019 ones, while the SDSS 2004 ($<9$\%) and TNG/VHR-R ($<6$\%) are too small to be appreciated.}
    \label{fig:bpt}
\end{figure*}

In Fig. \ref{fig:bpt} we show the [\ion{N}{ii}]- (left panel) and [\ion{S}{ii}]-BPT diagrams (right panel) with the line-flux ratio measurements obtained for the 4 available optical spectra: the archival SDSS data, and our TNG spectra acquired in 2019 (LR-B) and 2022 (LR-B and VHR-R). The VHR-R 2022 points combine the measurements of \textit{x}- and \textit{y}-axis ratios from the VHR-R and LR-B 2022 spectra, respectively. For clarity of the plots, we only represent the errorbars on the TNG/LR-B 2022 measurements, which are compatible with the LR-B 2019 ones, whereas the SDSS 2004 ($<9$\%) and TNG/VHR-R 2022 ($<6$\%) uncertainties on the flux ratios are too small to be appreciated in the two diagrams. The solid and dashed black lines are defined in \citet{2001ApJ...556..121K, 2006MNRAS.372..961K} and \citet{2003MNRAS.346.1055K}, and separate regions in the diagrams dominated by a different source of ionisation.

All our flux-ratio measurements confirm the previous classification of NGC 4156 as a Seyfert/LINER galaxy \citep{2016MNRAS.455.2551N}, additionally showing that no significant change in narrow-line flux ratios has occurred. In the [\ion{N}{ii}]-diagram, all the points lie in the AGN region but they are still consistent within the uncertainty with the composite region, where ionisation can be due to either star formation (i.e. HII regions) or an AGN, or a combination thereof.
The situation is very similar in the [\ion{S}{ii}]- BPT diagram, with all the 4 measurements compatible with the previous LINER classification, but also consistent with an AGN (Seyfert) or star formation as primary ionising source. As a consequence of this inconclusive BPT-based classification, definitive optical evidence of the AGN residing in NGC 4156 cannot be obtained from the narrow-line emission. In fact, the appearance of broad \ha\ and \hb\ components represents the first clear optical signature of the elusive AGN hosted by NGC 4156.

\end{appendix}
\listofobjects
\end{document}